\documentclass[12pt]{article}
\usepackage{graphicx}

\newcommand\pubnumber{}
\newcommand\pubdate{\today}

\def\institute{Department of Physics and Astronomy\\
University of Notre Dame, Notre Dame, IN 46556, USA}

\def\Title#1{\begin{center} {\Large #1 } \end{center}}
\def\Author#1{\begin{center}{ \sc #1} \end{center}}
\def\Address#1{\begin{center}{ \it #1} \end{center}}

\newcommand\pubblock{\rightline{\begin{tabular}{l} \pubnumber\\
         \pubdate  \end{tabular}}}
\newenvironment{Abstract}{\begin{quotation}  }{\end{quotation}}
\newenvironment{Presented}{\begin{quotation} \begin{center} 
             PRESENTED AT\end{center}\bigskip 
      \begin{center}\begin{large}}{\end{large}\end{center} \end{quotation}}

\newcommand{\tth}{\mathrm{t\bar{t}H}}
\newcommand{\ttv}{\mathrm{t\bar{t}V}}
\newcommand{\ttw}{\mathrm{t\bar{t}W}}
\newcommand{\ttbar}{\mathrm{t\bar{t}}}

\begin{document}
\begin{titlepage}
\pubblock

\vfill
\Title{Searches for $\tth$ and tHq with H$\rightarrow$leptons}
\vfill
\Author{Charles Mueller\\on behalf of the ATLAS and CMS collaborations}
\Address{\institute}
\vfill
\begin{Abstract}
The latest results from searches for a Standard Model Higgs boson produced
in association with a top quark pair ($\tth$) and single top quark (tHq) decaying
to final states with multiple leptons are presented using datasets from the CMS
and ATLAS experiments.
\end{Abstract}
\vfill
\begin{Presented}
$9^{th}$ International Workshop on Top Quark Physics\\
Olomouc, Czech Republic,  September 19--23, 2016
\end{Presented}
\vfill
\end{titlepage}
\def\thefootnote{\fnsymbol{footnote}}
\setcounter{footnote}{0}

\section{Introduction}
While the discoveries of the top quark and Higgs boson were essential in
verifying the standard model (SM) of particle physics, many important questions
remain unanswered. Specifically, why SM particles have the masses that are observed,
and whether or not the top quark's large mass comes only from its interaction with the Higgs.
The top quark is the heaviest fundamental particle, suggesting
it could play a special role in electroweak symmetry breaking. According to the SM, the top quark's large mass comes
from its Yukawa interaction with the Higgs. This property of the top quark
makes $\tth$ and tHq processes an excellent place to test the SM, while remaining
sensitive to new physics beyond the SM. By comparing the $\tth$ coupling with SM Higgs
production via a gluon-gluon loop with virtual top exchange, limits can be set on
new physics in the gluon-gluon loop. 

The analyses presented here cover recent 2016 LHC Run II $\tth$ measurements,
as well as previous tHq results by both the CMS~\cite{cms-jinst} and ATLAS~\cite{atlas-jinst} collaborations.
The Higgs decays targeted include
WW*, ZZ*, and $\tau\tau$. The Higgs and top systems may decay semi-leptonically or fully
leptonically, but the final state leptons must originate from both the top and Higgs systems.
The experimental signatures include two same-sign leptons and greater than or equal to three leptons, where leptons are defined as
muons or electrons. 
These requirements narrow the available phase space, which makes
signal processes rare, but eliminates large backgrounds.  

\section{$\tth\rightarrow$leptons}

\subsection{Object and event selection}

The CMS~\cite{cms-hig-16-022} and ATLAS~\cite{atlas-conf-2016-058} $\tth$ analyses begin with a similar object selection strategy. The most important aspect of
the object selection is the leptons.
The lepton selection uses 3 distinct classes of increasingly tight categories. The first category
of leptons is the preselection, consisting of cuts on kinematics, isolation, vertexing and experiment-specific identification
variables. The next class is the fakeable selection, which is enriched in non-prompt leptons originating from b-quark decays.
These leptons are used later for defining control regions. The final lepton category, known as the ``tight'' selection is used to select
prompt leptons for the signal region definitions. This selection includes tightened versions of the preselection cuts, and
additionally, for CMS only, a cut on a multivariate analysis (MVA) value used to identify prompt from non-prompt leptons. This is the primary difference between the ATLAS
and CMS tight lepton selections; instead of a cut on a MVA value, ATLAS uses tighter cuts on kinematics and isolation.
The remaining objects include jets, taus, and missing transverse energy (MET). These object definitions and selections are
similar between CMS and ATLAS. 

The event selections are also similar between ATLAS and CMS, and are primarily defined by the tight lepton multiplicity. The
selection with the highest statistics is the two-lepton same-sign selection with no hadronically decaying taus. In addition
to the nominal requirements, this selection requires lepton transverse momenta be greater than 25 GeV and the presence of
at least 4, 5 jets for CMS, ATLAS analyses respectively. The next selection is identical to the previous, but requires exactly
1 hadronic tau. Next are the three-lepton and four-lepton categories. While the ATLAS analysis uses each of these, CMS combines
them into a greater than or equal to three-lepton category. Here, the sum of the three lepton electric charges must be $\pm$1,
and there must be greater than or equal to 2, 3 jets for CMS, ATLAS respectively. Finally, a veto on the presence of two same-flavor opposite-sign
leptons within 10 GeV of the Z boson mass is required. An additional requirement for all categories is the presence of b-quark
jets. For CMS, this means at least 2 jets passing the loose working point of the b-tagging MVA discriminant, or at least one jet
passing the medium working point. For ATLAS, this corresponds to at least one jet passing the medium working point of a b-tagging
discriminant.

\subsection{Signal extraction and background estimation}
While the object and event selections are comparable between ATLAS and CMS, the signal extraction approaches differ significantly.
With tighter object definitions and more selective signal regions resulting in a higher signal purity, ATLAS employs a cut and count
approach in each of the six categories described in the previous section. 

The signal extraction strategy for CMS relies on a two-dimensional MVA technique. There are two boosted decision tree (BDT) MVAs,
each trained to discriminate against a single background. The backgrounds targeted individually are the two largest backgrounds,
$\ttv$, where V is a vector boson, and the fake lepton background from $\ttbar$. The output from each BDT is plotted on
a separate axis, forming a two-dimensional shape, which is then binned, forming a one-dimensional shape in each category. In addition
to the categories described previously, the CMS analysis splits the catgories described previously by the sign of the sum of the lepton
charges, and by the presence (or absence) of two medium b-tagged jets, referred to as the b-tight and b-loose categories. This
additional categorization provides additonal discrimination power. The BDTs are also trained separately in two-lepton same-sign, and 
three-lepton categories. The inputs for these BDTs consist of lepton and jet kinematics, solid angles, and MET variables. In addition
to these variables, the outputs from a Matrix Element Method (MEM) are also used as inputs to the BDT trained against the
$\ttv$ background (for the three-lepton category only). The MEM output is calculated using a signal hypothesis of $\tth$,$H\rightarrow$WW
assuming three final state leptons. The background hypothesis is $\ttv$. 

In contrast to the signal extraction techniques, the background estimation is similar between the ATLAS and CMS analyses. The main
backgrounds can be classified as reducible and irreducible. The reducible backgrounds consist primarily of non-prompt leptons from
b-quark decays in semi-leptonic $\ttbar$, but also include prompt leptons with incorrectly measured electric charge. Most importantly,
the reducible backgrounds are estimated via data-driven control regions. The irreducible backgrounds are estimated from Monte Carlo (MC)
samples and consist of $\ttv$ and diboson processes. The irreducible backgrounds are referred to as such because the signal regions do
not explicitly veto events from these processes.

\subsection{Results}
Both CMS and ATLAS results were produced with the 2015 and 2016 13 TeV LHC Run II datasets and are summarized below in Figure~\ref{fig:tth_results}.
The ATLAS analysis analyzed 13.2 $fb^{-1}$ of data,
while the CMS analyzed 15.2 $fb^{-1}$ of data. CMS measured the best-fit signal strength, $\mu = 2.0^{+0.8}_{-0.7}$, corresponding to a significance
of $3.2\sigma$ under the background-only hypothesis. ATLAS measured the best-fit $\mu = 2.5^{+1.3}_{-1.1}$, corresponding to a background-only
significance of $2.2\sigma$. The leading systematic uncertainties in each analysis include the non-prompt background estimation, jet-vertex association
and pile-up modeling, luminosity, signal and background modeling, and the jet energy corrections. 

%%%%%%%%%%%%%%%%%%%%%%%%%%%%%%%%%%%%%%%%%%%%%%%%%%%%%%%%%%%%%%%%%%%%%%%%%
%%
%%   use this format to include an .eps figure into your paper
%%
\begin{figure}[htb]
\centering
\includegraphics[height=2.5in]{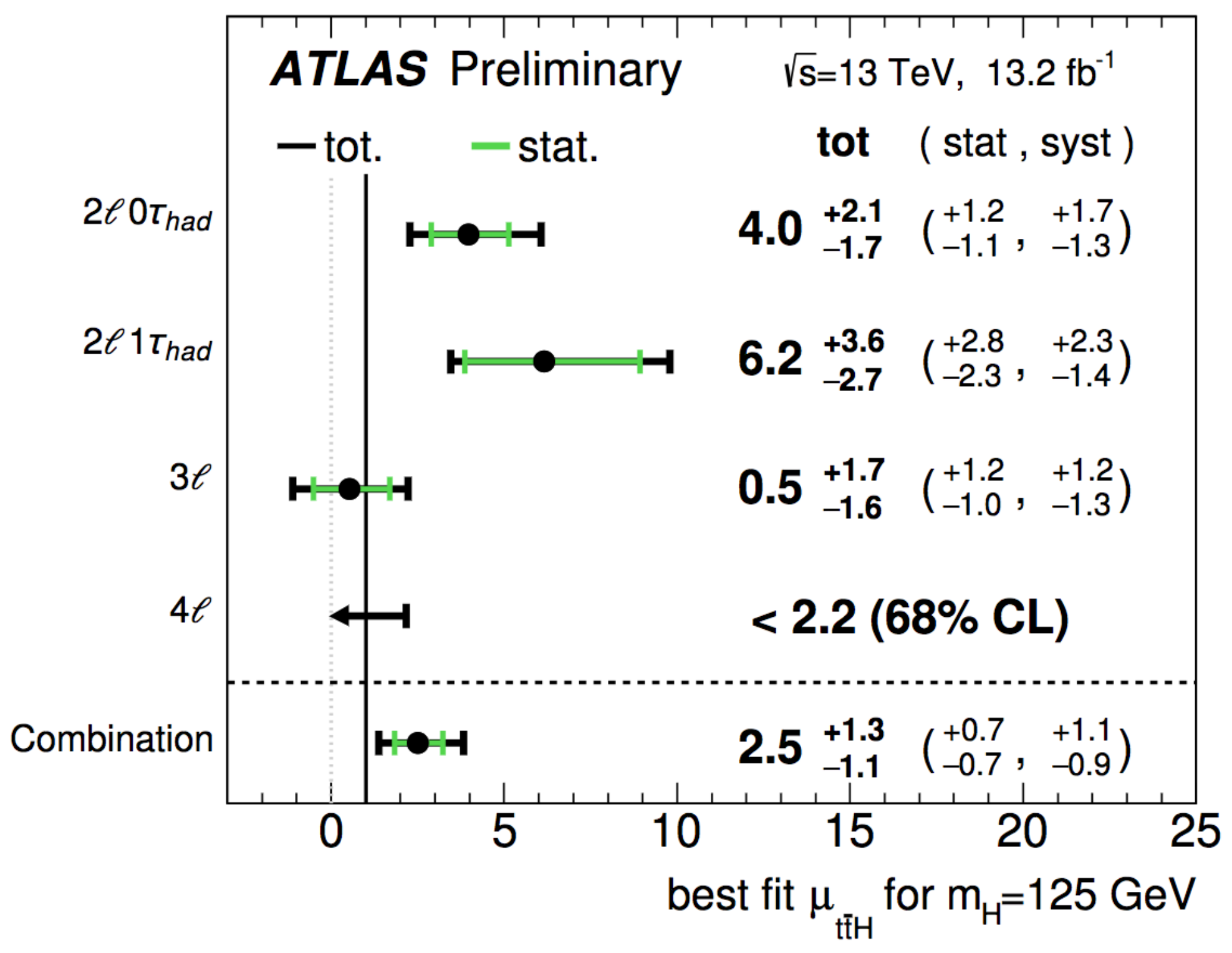}
\includegraphics[height=2.5in]{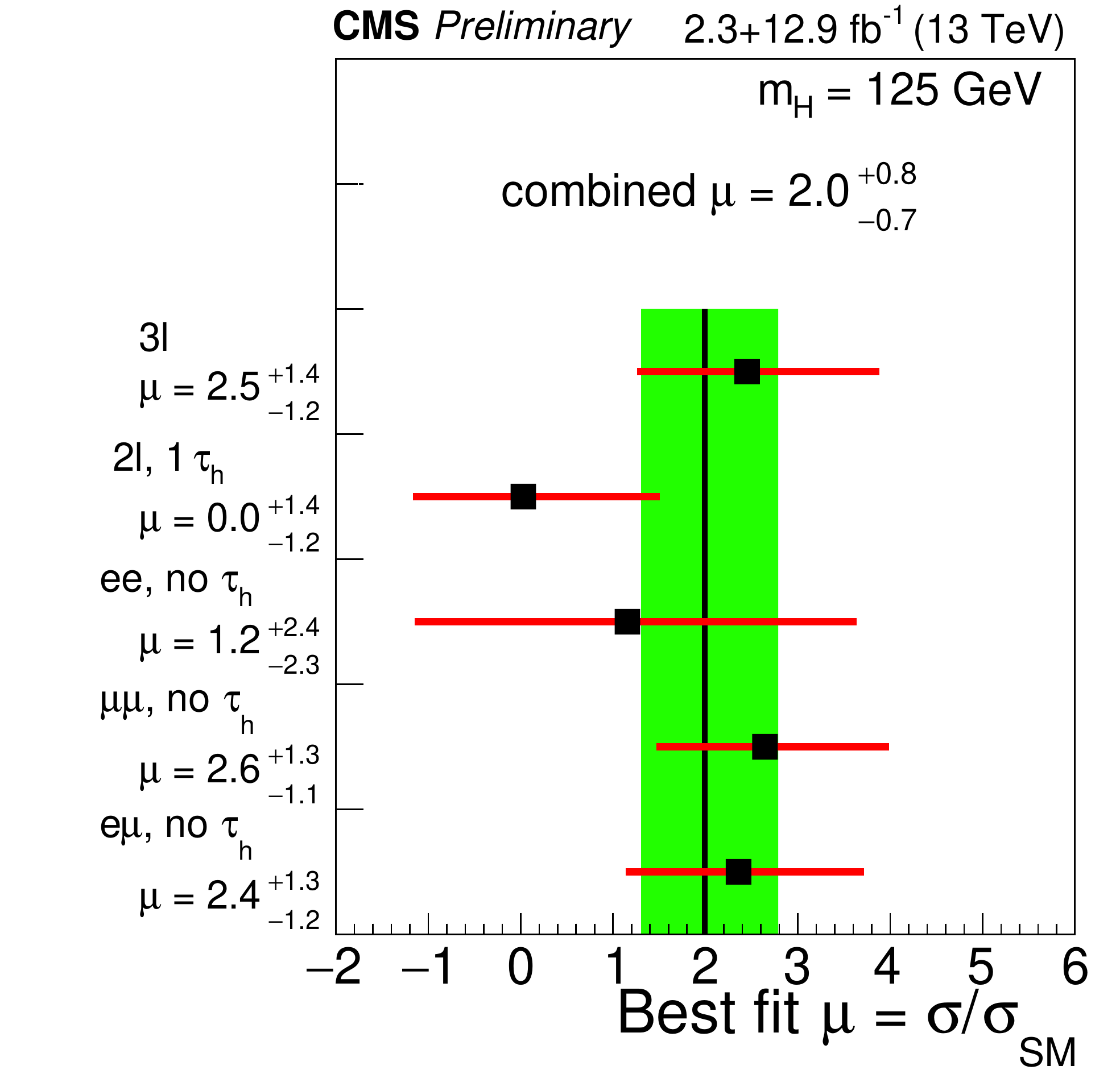}
\caption{A summary of the latest $\tth$ results for ATLAS (left) and CMS (right).}
\label{fig:tth_results}
\end{figure}
%%%%%%%%%%%%%%%%%%%%%%%%%%%%%%%%%%%%%%%%%%%%%%%%%%%%%%%%%%%%%%%%%%%%%%%%%%%

\section{$tHq$$\rightarrow$leptons}
The most recent tHq$\rightarrow$leptons measurement was made by CMS at 8 TeV~\cite{cms-hig-13-020}. This search targeted the WW and $\tau\tau$ decays of the Higgs, and measured
the interference between the two leading tHq production diagrams below in Figure~\ref{fig:thq_feynman}. This interference is highly sensitive to a non-SM inverted top-Higgs Yukawa coupling,
leading to a large enhancement in the tHq cross section under the inverted (negative) coupling scenario.

%%%%%%%%%%%%%%%%%%%%%%%%%%%%%%%%%%%%%%%%%%%%%%%%%%%%%%%%%%%%%%%%%%%%%%%%%
%%
%%   use this format to include an .eps figure into your paper
%%
\begin{figure}[htb]
\centering
\includegraphics[height=1.5in]{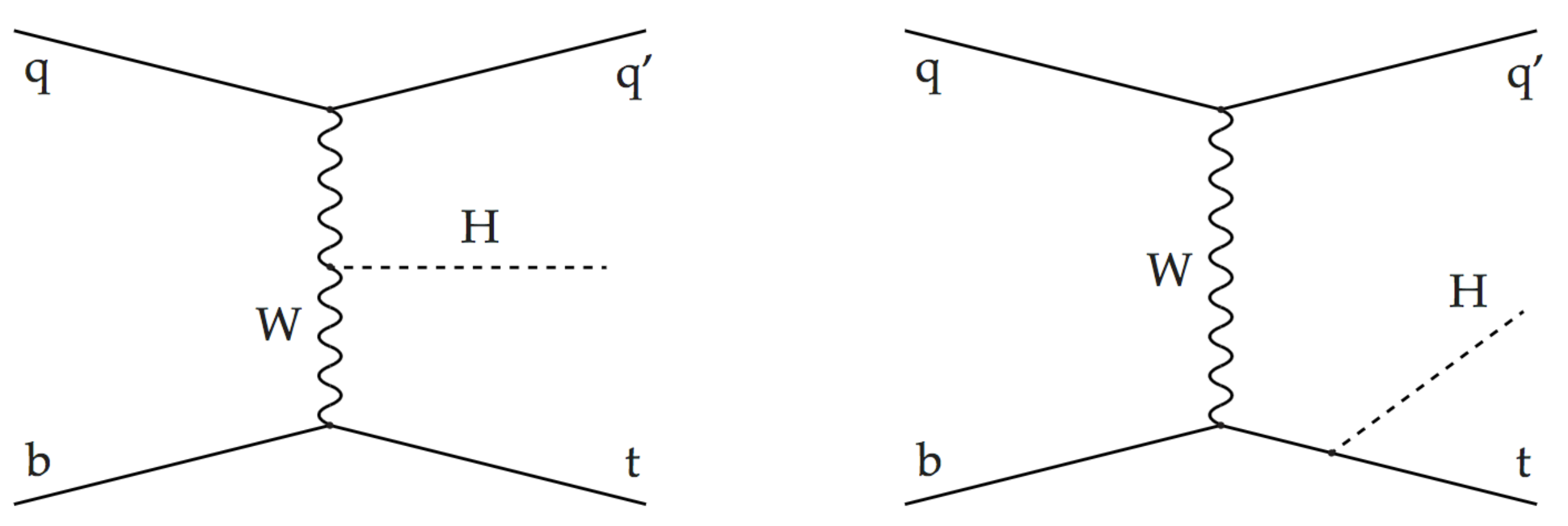}
\caption{The two interfering tHq process diagrams.}
\label{fig:thq_feynman}
\end{figure}
%%%%%%%%%%%%%%%%%%%%%%%%%%%%%%%%%%%%%%%%%%%%%%%%%%%%%%%%%%%%%%%%%%%%%%%%%%%

\subsection{Object and event selection}
The object and event selection use a similar strategy as the $\tth$ analysis detailed previously. Prompt leptons are identified with the same MVA
mentioned previously, while the other object selections are also very similar to those in the $\tth$ analysis.

The event selection consists of a two-lepton same-sign category and a three-lepton category. The two-lepton same-sign category requires lepton
transverse momenta greater than 20 GeV, no dilepton mass greater than 20 GeV, no hadronic taus, at least one central jet ($|\eta<1.0|$), at least one
central b-tagged (loose working point) jet, and at least one forward jet ($|\eta>1.0|$). The three-lepton category requires exactly three leptons with
transverse momenta greater than 20,10,10 GeV, the MET must be greater than 30 GeV, and a Z veto is required. In addition to these, all of the jet requirements
for the two-lepton same-sign category are required, but the b-tagging working point is tightened from loose to medium. This event selection strategy defines a
relatively loose selection, followed by a multivariate signal extraction. 

\subsection{Signal extraction and background estimation}
A linear likelihood discriminator separates signal from background. The signal extraction relies on three types of input variables: forward jet activity, which
is only present in the signal process, jet and b-jet multiplicity, and lepton kinematic and charge variables. Two separate likelihoods were used, one for each
event selection described previously. The likelihood was trained against loosened selections of $\ttbar$, WZ, WWqq, and $\ttv$ processes for the two-lepton same-sign
category. For the three-lepton category, the likelihood was trained against loosened selections of $\ttbar$ and $\ttv$.   

The background estimation in the tHq analysis uses the same approach as the $\tth$ analysis. The leading backgrounds to tHq are non-prompt (fake) leptons primarily from
semi-leptonic $\ttbar$ decays, leptons with incorrectly measured charge sign, and $\ttw$. The non-prompt and charge mis-measured backgrounds are estimated using
a data-driven technique via control regions. The remaining backgrounds are estimated from MC.  

\subsection{Results}
The tHq analysis places 95$\%$ C.L. upper limits on tHq production under the inverted top-Higgs coupling with 19.5 $fb^{-1}$ of 8 TeV data. The upper limit
was measured to be 9.3 ($8.1^{+6.0}_{-11.8}$) observed (expected) in the two-lepton same-sign $\mu\mu$ channel, 11.4 ($9.3^{+7.0}_{-13.5}$) observed (expected)
in the two-lepton same-sign $e\mu$ channel, 11.5 ($8.6^{+6.6}_{-12.4}$) observed (expected) in the three-lepton channel, and finally 6.7 ($5.0^{+3.6}_{-7.1}$)
observed (expected) for all channels combined. The leading systematics include the estimation of the non-prompt background, and the signal and background modeling. 

\section{Conclusions}
The $\tth$ and tHq analyses probe the top-Higgs Yukawa coupling at tree-level. The new 2016 $\tth$ multilepton results are compatible within approximately
$1\sigma$ of the SM with $m_{H}=125$ GeV. The ATLAS $\tth$ analysis measured a best-fit $\mu = 2.5^{+1.3}_{-1.1}$ corresponding to a significance of $2.2\sigma$.
The CMS $\tth$ analysis measured a best-fit $\mu = 2.0^{+0.8}_{-0.7}$ corresponding to a significance of $3.2\sigma$. The CMS tHq analysis set 95$\%$ C.L. upper limits
on the production of the tHq process under the inverted top-Higgs coupling scenario of 6.7 ($5.0^{+3.6}_{-7.1}$) observed (expected). These results spell exciting
prospects for analyses on the full 2016 dataset and beyond. 

%\Acknowledgements

\end{document}